\documentclass[twocolumn,showpacs,amsmath,amssymb]{revtex4}
\usepackage{graphicx}
\usepackage{dcolumn}
\usepackage{amsmath}

\newcommand{\lessim} {\mathop{\,<\kern - 1.05 em \lower 1.ex \hbox {$\sim$}\,}}
\newcommand{\grtsim} {\mathop {\,> \kern - 1.05 em \lower 1.ex \hbox 
{$\sim$}\,}}

\begin{document}

\normalsize  \title{
Magnetic oscillations and
field induced spin density waves in (TMTSF)$_2$ClO$_4$}

\author{Danko Radi\'c, Aleksa Bjeli\v{s}}
\affiliation{Department of Physics, Faculty of Science, University of
Zagreb, POB 162, 10001
Zagreb, Croatia}
\author{Dra\v{z}en Zanchi} 
\affiliation{Laboratoire de Physique Th\'eorique et Hautes Energies,
Paris, France.}

\begin{abstract}

We report an analysis of the effects of magnetic field on a quasi-one-dimensional
band of interacting electrons with a transverse dimerizing potential. One-particle
problem in bond-antibond representation is solved exactly. The resulting propagator
is used to calculate the spin-density-wave (SDW) response of the interacting system
within the matrix RPA for the SDW susceptibility. We predict the magnetic field
induced transition of the first order between interband SDW$_0$ and intraband
SDW$_{\pm}$ phases. We reproduce the rapid oscillations with a period of 260 Tesla and 
the overal profile of the TMTSF$_2$ClO$_4$ phase diagram.


\end{abstract}

LPTHE/02-33 
\bigskip

\maketitle

\bigskip



\newpage

Investigations of  quasi-one-dimensional electronic systems at high 
magnetic fields  and at low temperatures continue to give an important insight 
into the one-particle properties and interaction-induced phases such as 
spin- and charge-density-wave, superconductivity, and Mott localization.
 \cite{review98}
One of most spectacular phases  of this kind are 
field-induced spin density wave (FISDW), found in Bechgaard salts
\cite{Chaikin}  and in some other low-dimensional compounds.\cite{Biskup99}
The phenomenon of the FISDW is well understood in the Bechgaard salt 
(TMTSF)$_2$PF$_6$ where the cascade of SDW phases with quantized 
 wave-vector is induced
by orbital effects of magnetic field to the quasi-one-dimensional 
orbits of band electrons. Theory based on the mechanism of quantized nesting 
\cite{QNM} reproduces satisfactorily main experimental data for this salt.

In this letter we concentrate on (TMTSF)$_2$ClO$_4$, a Bechgaard salt 
which after a slow cooling \cite{Qualls00,Matsunaga} enters into a qualitatively different type
of FISDW phase at low temperatures, with a  phase diagram that is still,
after more than ten years of intensive studies \cite{Chaikin,review98}, a matter
of both experimental and theoretic controversies. In particular for 
magnetic field B$>$8T  the nature of the ordering in
the relaxed material   is not 
a simple  FISDW with some low integer quantum number N, 
but a qualitatively diferent 
state containing  several puzzling subphases.\cite{Chaikin,McKernan95,Chung00}
This phase is at 8T separated by a line of 
first order transition from a cascade of FISDW phases which very much resembles
to that of the standard model. Another characteristic phenomenon, the 
rapid oscillations (RO) in $1/B$ with a
frequency of 260 Tesla are visible 
in transport properties  in both metallic  {\em
and} FISDW state.\cite{Chung00,Kang01,Chaikin} Similar RO are seen also in thermodynamic quantities like
torque, magnetization, sound velocity and specific heat, but only in the ordered
phase.\cite{review98,Chaikin}
The highest value of $T_c$ in the 
$T_c(B)$ dependence is 5.5K, instead of 12K as expected from analogy with
the (TMTSF)$_2$PF$_6$ salt.

The incompatibility of 
above facts with the quantum nesting model (QNM) for a single quasi-1D band
is believed to stem from the particular ordering of ClO$_4$ anions.\cite{Chaikin}
This ordering introduces the new modulation with the 
wave vector $(0,\pi/b, 0)$, i. e. a dimerization in the low-conducting 
direction with the inter-chain distance $b$. 
The magnitude of the dimerizing potential can be tuned to some extent by 
varying the cooling rate \cite{Qualls00,Matsunaga}. Thus, anions presumably 
remain disordered in the rapidly quenched samples. Then there is no 
dimerization gap in the band, and the system shows 
properties of a {\em single} 
quasi-1D imperfectly nested band with a SDW order appearing already in the 
zero magnetic field \cite{Qualls00,Matsunaga,Bjelis-Maki92}. The anion
ordering in slowly relaxed samples is at about 24K, and coincides with
the onset of
rapid oscillations in the magnetoresistance.\cite{Uji97}

The dimerized band has two pairs of Fermi sheets in the new Brillouin zone. Already
simple geometric arguments \cite{McKernan95} suggest three possible nesting wave
vectors favoring various SDW phases.  First, interband nesting, leads to SDW$_0$
that is the two-band version of the standard FISDW phase. Other two nesting vectors
relate Fermi sheets within the same band. They give SDW$_+$ for antibond nesting and
SDW$_-$ for bond nesting. However the interplay between SDW$_0$ and SDW$_{\pm}$ is
not only a geometric question of the choice of the nesting vector. Due to a finite
anion potential $V$ in the kinetic part of the Hamiltonian an off-diagonal term
appears in the SDW response, making necessary an appropriate matrix approach
\cite{ZB01,Dupuis02} in the calculation of the  critical
susceptibilities. The response matrix is formulated in
the space of two order parameters, $\Delta _h$ (``homogeneous'') and $\Delta _a$
(``alternating''), determining the magnetic pattern
\begin{equation} \label{mag}
m_z(x,R_{\perp})=
(\Delta _h \pm  \Delta _a)\cos{\left[ (2k_F+k)x+
pnd \right]} \; .
\end{equation}
Here $d\equiv2b$ and the upper and lower sign stay for even $(R_{\perp}=nd)$ and
odd $(R_{\perp}=nd+d/2)$ chains respectively. As it is shown in Refs.
\cite{ZB01,Dupuis02}, SDW$_0$ [$\Delta _h \neq 0,  \Delta _a =0$] is stabilized for
low values of $V$ (providing the imperfect nesting parameter $t'_{b}$ allows for SDW
stabilization), while SDW$_{\pm}$ with $\Delta _h \geq \Delta _a  \neq 0$ gets
stable for $V/t_{b} > 1.6$ irrespectively to the value of $t'_{b}$. Here $t_{b}$ is
the interchain hopping integral. The slowly relaxed (TMTSF)$_2$ClO$_4$ samples are
expected to lie in the range of intermediate values of $V$ in which there is no SDW
ordering at $B=0$ down to $T=0$.

Indeed, as it will be shown, $V/t_b$ fitting the
experiments is close to one.
In this range it is not allowed to use the quasi-classical approximation of Gor'kov
and Lebed \cite{Gorkov}, which consists in making Peierls substitution $p\rightarrow
p-eA$ in each sub-band separately and including the anions' effects only via
magnetic breakdown (MB) junctions near the zone boundary. While this approximation
is sufficient for $V/t_b\ll 1$, here one has to solve the whole quantum-mechanical
problem instead.

\begin{figure*}[htbp]
\begin{center}
\setlength{\unitlength}{1cm}
\begin{picture}(9,10.5)
  \put(-2.5,-0.5){\includegraphics[width=12cm]{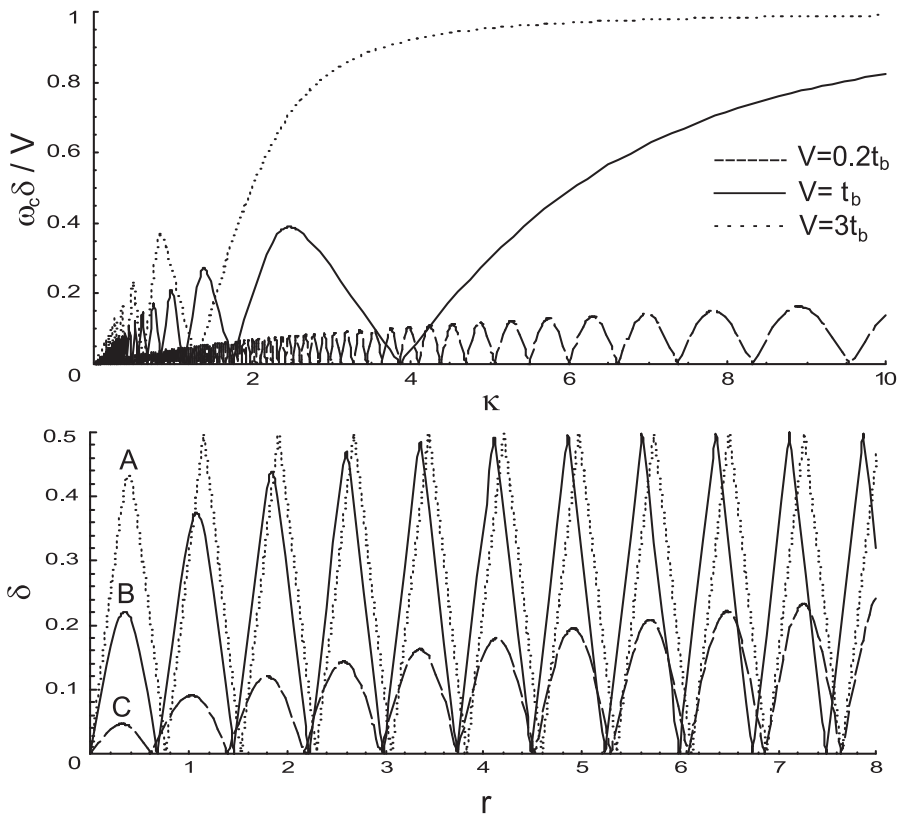}}
\end{picture}
\caption{(a) Energy ratio $\omega_c \delta/V$ as function of the magnetic
breakdown parameter $\kappa$ for several values of $V/t_b$. (b) Dependence of
$\delta$ on $r$ for  $\theta=10^{\circ}$ (A), $45^{\circ}$ (B), 
and $80^{\circ}$ (C). }
\label{Floquet}
\end{center}
\end{figure*}
\begin{figure*}[htbp]
  \begin{center}
    \setlength{\unitlength}{1cm}
    \begin{picture}(17,12)
       \put(0,-0.5){\includegraphics[width=17cm]{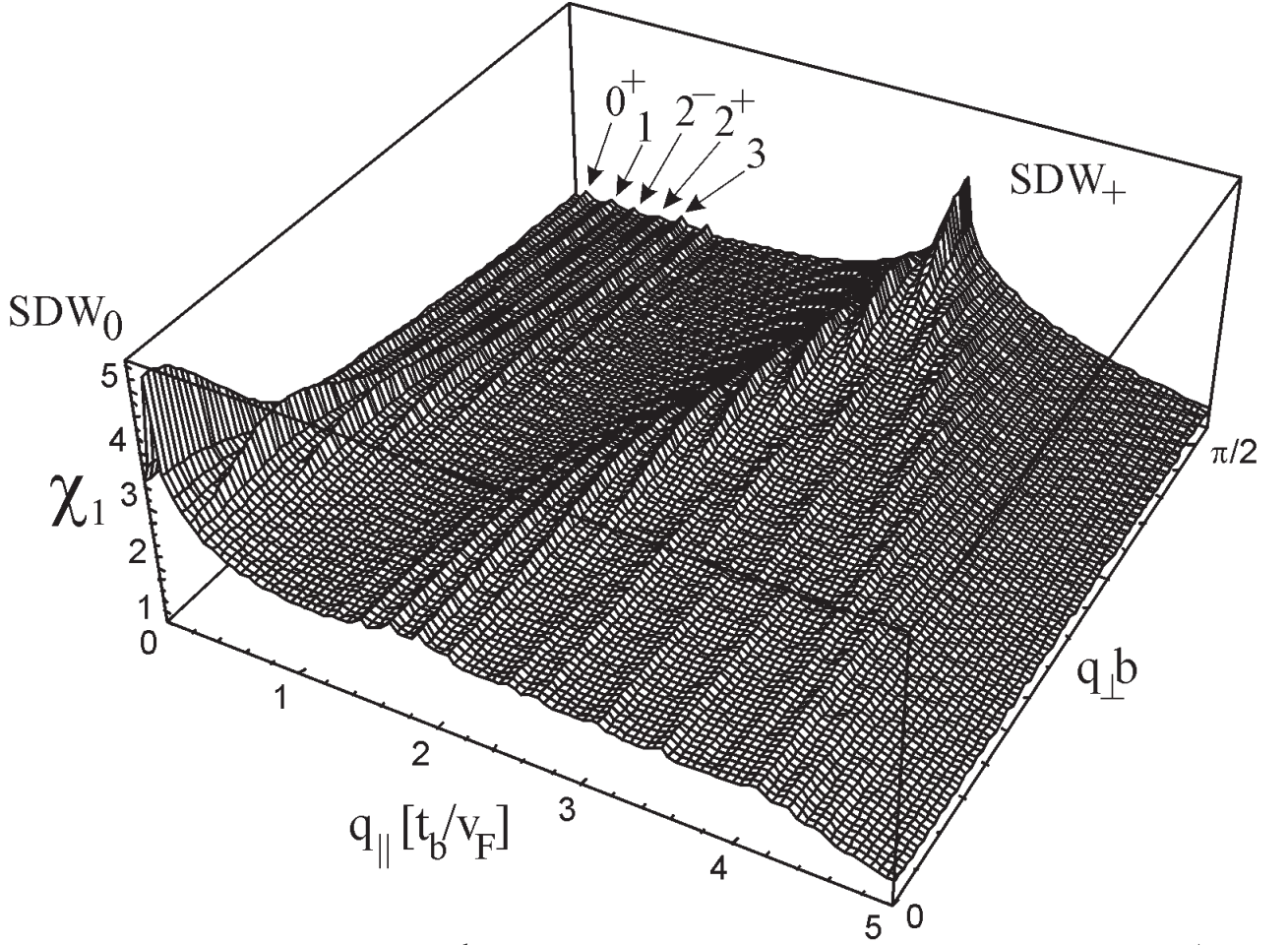}}
    \end{picture}
    \caption{Susceptibility $\chi_1$ [in units of $(2\pi v_F)^{-1}$]. 
     Arrows indicate the longitudinal
    coorinates of the peaks at $ 2\delta$ $(0^+)$, 
$G$ $(1)$, $2G-2\delta$ $(2^-)$, 
$2G+2\delta$ $(2^+)$, and $3G$ (3). Maximum
    of $\chi_1$ corresponds to the phase $(0^+)$ for ${\bf q}=(2G,0)$.}
    \label{chi}
  \end{center}
\end{figure*}
\begin{figure*}[htbp]
  \begin{center}
    \setlength{\unitlength}{1cm}
    \begin{picture}(17,11)
      \put(-1,-1){\includegraphics[width=17cm]{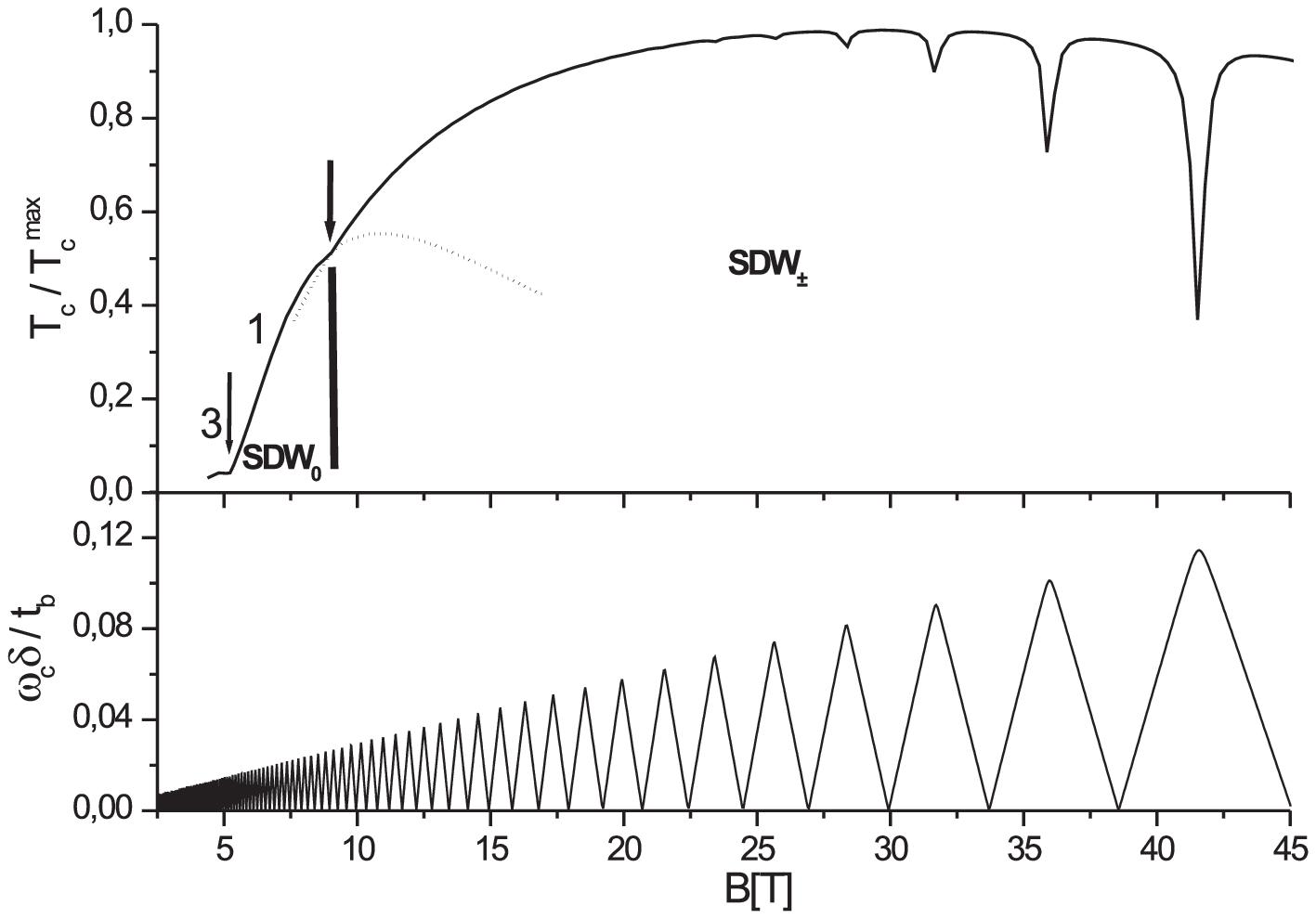}}
    \end{picture}
    \caption{(a) Phase diagram. (b) Energy ratio $\omega_c \delta/t_b$
     on the same magnetic scale as the phase diagram.}
    \label{fd}
  \end{center}
\end{figure*}
It was pointed out several times (see \cite{Chung00} and references therein) that a
mechanism of coherent inter-band tunnelling, very similar to Stark over-gap quantum
interference (QI) in magnesium \cite{Stark74}, is essential for high-field physics
in (TMTSF)$_2$ClO$_4$. In particular, RO in metallic state can be explained only in
terms of QI mechanism because no closed orbits exist. On the contrary, in the SDW
state both closed orbits {\em and} Stark interference contribute to RO. Oscillating
behavior periodic in $1/B$ can be seen already at
the level of one-particle spectrum. Namely, electronic propagator with longitudinal
momentum $k$ has poles at
\begin{equation} \label{spectrum}
E_f= v_F\left[ f(k-k_F) +GN\right]  \pm v_FG\delta \; ,
\end{equation}
where $f$ is left-right index,  $G\equiv eBb/\hbar$ is the magnetic
wavenumber and $N$ is integer number. The first term in eq.(\ref{spectrum})  
is the standard QNM dispersion and the last term is the splitting due to
anions. Over-gap resonances are present in $\delta(B)$ as will be 
discussed below (see Fig.\ref{Floquet}).
The
expression for the  spectrum (\ref{spectrum}) is common to perturbation calculations
\cite{Osada92}, to quasi-classical tunnelling analysis \cite{Gorkov}, and to our
exact solution as well.
What changes from one approach to another are the dependence $\delta(B)$ and
the result for electronic wavefunction. In order to obtain them exactly
we start from the effective one-particle Hamiltonian for electronic
operators $\Psi_f(x,p)$
 \begin{equation} 
H_0=   iv_F\rho_3  \partial _x +  
 \tau_3 {\cal T}(pb-Gx)+\tilde{\cal T}(pb-Gx)
-V\tau _1
\; ,
\label{hkin}
\end{equation}
where $\rho$'s and $\tau$'s are Pauli matrices in 
left-right and bond-antibond indices respectively.
The most general transverse dispersion was split into two parts
 \begin{equation} 
{\cal T}(pb)\equiv 2 \sum_{j=1}^{\infty}t_j \cos[(2j-1) pb 
]\; ,
\tilde{\cal T}(pb)\equiv 2 \sum_{j=1}^{\infty}t'_j \cos [ 2j pb ]
\label{disp}
\end{equation}
corresponding to effective hoppings \cite{Yamaji82}
between odd and even neighbors respectively. 
We diagonalize $H_0$ by the unitary transform 
 \begin{equation}
\Psi _f= 
\left(
\begin{array}{cccc}
\alpha _f & \beta_f \\
-\beta _f^* & \alpha ^*_f \\
\end{array}
\right) e^{if\theta} \Phi _f \; ,
\label{transf}
\end{equation}
with $|\alpha|^2+|\beta|^2=1$, and functions $\alpha$, $\beta$ and $\theta$
depending on $x$ and $p$ only through the combination $z=pb-Gx$.
 From the
requirement that the effective Hamiltonian for field $\Phi$ is only
$ifv_F\partial_x$ we get
\begin{equation}
\theta(z)=\frac{1}{v_F}\int ^z dz \, \tilde{\cal T}(z)
\label{theta}
\end{equation}
and a system of differential equations for functions $\alpha$ and $\beta$
\begin{equation}
\begin{array}{cc}
ifv_F\alpha '_f(z)=-{\cal T}(z)\alpha _f(z)-V\beta_f^*(z)\\
ifv_F\beta '_f(z)=-{\cal T}(z)\beta _f(z) +V\alpha _f^*(z).
\end{array}
\label{system}
\end{equation}
Note that $\theta(z+2\pi)=\theta(z)$ and that 
$\alpha_+(z)=\alpha_-^*(z)$ and  $\beta_+(z)=\beta_-^*(z)$, so that it suffices
to follow e. g. solutions $\alpha _+(z), \beta _+(z)$ of the system (\ref{system}).
According to Floquet theory these solutions can be written in the form
$\alpha(z)=A(z)\exp({-iz\delta })$; $\beta(z)=B(z)\exp({iz\delta })$. $A$ and $B$
are periodic with the period $2\pi$, and the closer inspection shows that the
Floquet exponent $\delta$ for the system (\ref{system}) is real for all values of
parameters, at least after keeping in ${\cal T}(z)$ only the leading term
$t_1$.

Once we find $A$, $B$ and $\delta$ the wave functions 
$\langle x,p|F_k\rangle$ of the states created by $\Psi ^{\dagger}(x,p)$ are 
known. The corresponding spectrum is one-dimensional,
$E_k=v_Ff(k-k_F)$. Projection of $|F_k\rangle$  to the
plane wave $\langle k',p|$ is
\begin{eqnarray}
\nonumber
\lefteqn{\sum_N{\biggl{\{ }} \mu_ke^{i(N-\delta)pb}\delta[k'-k+G(N-\delta)]
\left(
\begin{array}{cc}
a_N\\
-\hat{b}_N
\end{array}
\right) + }& \\
& + & \nu_ke^{i(N+\delta)pb}\delta[k'-k+G(N+\delta)]
\left(
\begin{array}{cc}
b_N\\
\hat{a}_N
\end{array}
\right)
{\biggr{\}} }
\label{decomp}
\end{eqnarray}
Coefficients $a_N$, $b_N$, $\hat{a}_N$ and $\hat{b}_N$ are Fourier components of the
products $A\exp({i\theta})$, $B\exp({i\theta})$, $A^*\exp({i\theta})$ and
$B^*\exp({i\theta})$ respectively. 
Coefficients $\mu _k$ and $\nu _k$ are fixed by initial conditions.
The expression (\ref{decomp}) tells us how the
plane wave $\exp({ikx+pR_{\perp}})$ 
is decomposed into discrete states $N$. Each
state $N$ is split by  $\delta$ in a way that components with 
the tilt $-\delta$
have the statistical weight $|\mu_k|^2(|a_N|^2+|\hat{b}_N|^2)$ 
and the ones with the tilt
$+\delta$ have the weight $|\nu_k|^2(|b_N|^2+|\hat{a}_N|^2)$. Green function $\langle\Psi (x,p)
\Psi^{\dagger}(x',p)\rangle$ is easily constructed using transformation
(\ref{transf}) and knowing that
$\langle\Phi\Phi^{\dagger}\rangle=(i\omega_n-iv_Ff\partial_x)^{-1}=G_{1D}$.

The Floquet exponent $\delta$ and the functions $A$ and $B$ are calculated using the
Hill's theory and the fundamental matrix method \cite{future,eqdif}. In the present
work we limit our calculations to first harmonics in Eq. \ref{disp} only,
parameterized with $t_1=t_b$ and $t_1'=t_b'$. Let us concentrate on the magnetic
field dependence of the Floquet exponent $\delta$ that splits the QNM spectrum as
given by Eq. \ref{spectrum}. Figure \ref{Floquet}(a) shows the energy
$\omega_c \delta$ (in units of $V$)
as a function of the magnetic breakdown parameter $\kappa \equiv 2\omega_ct_b/V^2$,
where $\omega_c=v_FG$ is the cyclotron frequency. In quasi-classical picture
$\kappa$ determines the probability of the over-gap tunnelling
$P=\exp({-\pi/2\kappa})$.\cite{Gorkov} One sees that the crossover from oscillating to saturating
behavior does not coincide with the crossover from the weak ($\kappa<1$) to the
strong ($\kappa>1$) MB. The position of the last zero of $\delta$ is not universal
in $\kappa$, but approximately in $r\equiv [(\gamma V)^2 + t_b^2]^{1/2}/\omega_c$,
where the value of $\gamma$ is $0.77$. Fig. \ref{Floquet}(b) shows $\delta (r)$
for several "polar angles" defined by $\tan \theta \equiv t_b/\gamma V$.
Oscillations of $\delta$ are approximately periodic in $r$ with a period of 
$0.80$. Choosing the parameters $t_b= 300K$, $v_F= 2\times 10^5 m/s$,
and $b = 7.7\times 10^{-10}m$  we fit RO at 260
Tesla by putting $V \approx 0.8 t_b$.

Taking the limit of strong magnetic filed $\omega_c/t_b\gg 1$ and of weak anion
potential $V/t_b \ll 1$ we can easily reproduce the 1D spectrum of Osada et al.
\cite{Osada92}, $E_k\rightarrow fv_F(k-k_f)\pm \omega_c \delta$ with $\delta
\rightarrow (V/\omega_c){\cal J}_0(4t_b/\omega_c)$, ${\cal J}_0$ being the
Bessel function. On the other hand the spectrum
of Gor'kov and Lebed \cite{Gorkov} is reproduced for weak
anion potential, $V/t_b \ll 1$. \cite{future}  The above fit, as well as
other insights \cite{Yoshino99,Lepeleven}  however strongly suggest
that $V$ in (TMTSF)$_2$ClO$_4$ is rather large, i. e. comparable to $t_b$.

We proceed with the solution of the interacting problem. Neglecting the
absence of a presumably small Umklapp scattering, the effective coupling for SDW is
the forward scattering amplitude $g_2$, here simply denoted by $U$. We employ the
matrix RPA formalism developed in Ref.\cite{ZB01}.  The resulting relevant
 bare susceptibility  is
$\chi_1({\bf q};T)=\frac{1}{2}{\{ }\chi_{aa}+\chi_{hh}+[ \left(\chi_{aa}-
\chi_{hh}\right) ^2+ 4(\chi_{ha})^2]^{1/2} {\} } $, playing in
the Stoner criterion
\begin{equation}
1-U{\chi}_1({\bf q}_c,T_c)=0, 
\label{Stoner}
\end{equation}
${\bf q}_c$ being the wave vector at which ${\chi}_1({\bf q})$ has the
maximum. 
The ratio of two SDW order parameters from
Eq.(\ref{mag}) is also a function of bare correlators
$\chi_{aa},\chi_{hh},\chi_{ah}$ in the $(a,h)$ basis (see \cite{ZB01}). The bare
correlators in the magnetic field are given by
\begin{align}
\nonumber
  \chi_{hh}  &=  \sum _N\left[ |I_{h0}|^2P_0+\frac{1}{2} I_{h+}^2P_+
+\frac{1}{2} I_{h-}^2P_-\right]   \; ,  \\ 
 \chi_{aa}  &=  \sum _N\left[ |I_{a0}|^2P_0+\frac{1}{2} I_{a+}^2P_+
+\frac{1}{2} I_{a-}^2P_-\right]   \; ,  \nonumber\\
 \chi_{ha}  &= \sum _N\left[\Re (I_{h0}I_{a0}^*)P_0+
\frac{1}{2} I_{h+}I_{a+}P_+-\frac{1}{2} I_{h-}I_{a-}P_-\right]\; , 
\label{haha}
\end{align}
where $P_0,P_{\pm}$ stand for $P(q_{\parallel}-NG,T)$ and
$P[q_{\parallel}-G(N\pm2\delta),T]$ respectively, 
$P(k,T)$ being the familiar 1D
Lindhard function at the wave number $2k_F+k$. $P_0$ and $P_{\pm}$ are the
inter-band and the intra-band susceptibilities of the $N$-th split level in the
decomposition (\ref{decomp}). The dependence on the transverse momentum is present
in the amplitudes $I(q_{\perp},N)$,
\begin{align}
I_{h0}(q_{\perp},N)&=\sum _n\left( a_nb_{N-n}-\hat{b}_n\hat{a}_{N-n}\right)e^{i
(n-N/2)q_{\perp}}\nonumber \\
I_{h+}(q_{\perp},N)&=\sum _n\left(
\hat{a}_n\hat{a}_{N-n}+{b}_n{b}_{N-n}\right)e^{i(n-N/2)q_{\perp}}\nonumber\\
I_{h-}(q_{\perp},N)&=\sum _n\left( a_na_{N-n}+\hat{b}_n\hat{b}_{N-n}\right)e^{i
(n-N/2)q_{\perp}}\nonumber\\
I_{a0}(q_{\perp},N)&=\sum _n\left( a_n\hat{a}_{N-n}-\hat{b}_n{b}_{N-n}\right)e^{i
(n-N/2)q_{\perp}}\nonumber\\
I_{a+}(q_{\perp},N)&=\sum _n\left( \hat{a}_nb_{N-n}+{b}_n\hat{a}_{N-n}\right)e^{i
(n-N/2)q_{\perp}}\nonumber\\
I_{a-}(q_{\perp},N)&=\sum _n\left( a_n\hat{b}_{N-n}+\hat{b}_n{a}_{N-n}\right)e^{i
(n-N/2)q_{\perp}}
\label{Iovi}
\end{align}
There are two important selection rules for these amplitudes, namely
for $N$ even, 
$I_{h0}(N)=I_{a0}(N)=0 $ while for $N$ odd,
$I_{h\pm}(N)=I_{a\pm}(N)=0 $.
Thus the interband processes contribute only to FISDW phases with 
odd $N$ while the
intraband processes contribute only to phases with even $N$. 
Consequently only 
phases with even  $N$ "see" the splitting by $\delta$.

The ${\bf q}-$dependence of the susceptibility $\chi_1({\bf q})$ for the particular
choice of parameters, $\omega _c=0.1t_b, V=0.8t_b, t_b'/t_b=0.03, T/t_b=0.001$, is shown in Fig.
\ref{chi}. The overall envelope assumes the shape present already in the absence of
magnetic field \cite{ZB01}. It is now superimposed by a well known characteristic of
FISDW susceptibilities \cite{Chaikin}, logarithmic peaks corresponding to single
one-dimensional bubbles $P(k)$, weighted by $p$-dependent amplitudes as defined by
Eqs.(\ref{haha},\ref{Iovi}). Qualitatively new feature regarding these peaks is the
splitting of peaks with even $N$ by $\pm \delta$ around the positions at
$k=NG$.

According to Eq.(\ref{Stoner}) at $T=T_c$ the highest of peaks in Fig. \ref{chi}
attains the value $1/U$. Fig.\ref{fd} shows the resulting phase diagram for a
realistic choice of parameters, $V=0.85t_b$, $t_b'=0.03t_b$ and $T_c(V=t_b'=0)=13K$.
The resulting maximal critical temperature within the present field range is
$T_c^{\mbox{max}}\approx 1.1K$. The most obvious characteristic of the obtained
phase diagram is the first order transition from SDW$_0$ to SDW$_{\pm}$ at
$B_c\approx 9$ Tesla. Dependence $T_c(B)$ for $B<B_c$ is similar to the FISDW
cascade in TMTSF$_2$PF$_6$, with the difference that here only odd phases
appear because the even ones are suppressed by splitting.
 For $B>B_c$ the critical temperature increases towards
the highest value $T_c^{\mbox{max}}$. As the magnetic field further increases
the critical temperature $T_c(B)$ starts to oscillate, with the sharp dips corresponding
to commensurability condition $2G\delta=G$ between the Floquet wave number and the
magnetic wave number.

Note that
the present choice of anion potential $V=0.85t_b$ places us in the intermediate
regime on scale $V$ \cite{ZB01} where the response of both SDW$_0$ and SDW$_{\pm}$
is suppressed for $B=0$. Hence, being of comparable magnitude the two instabilities
compete once they are restored by magnetic field. Both phases in fig.\ref{fd} are
sensitive to $V$, general trend being that by increasing $V$ one decreases
$T_c(SDW_0)$ and increases $T_c(SDW_{\pm})$. The parameter of imperfect
nesting in the standard model, $t_b'$, here affects only SDW$_0$, while for
SDW$_{\pm}$ $t_b'$  plays a role of an effective nearest neighbor hopping. We
remind that the effective parameter of imperfect nesting for SDW$_{\pm}$ is a
function of $t_b/V$, as pointed out in Ref. \cite{ZB01}. $t_b'$ acts on SDW$_0$ in a
standard way \cite{QNM}, i. e. it fixes the width of the FISDW cascade $\Delta
\omega_c\sim t_b'$ so that by increasing $t_b'$ one reduces $T_c(SDW_0)$ at fixed
$B$.

The result of the subtle interplay between two scales $V$ and $t_b'$ is that
the realistic profile of the phase diagram is possible only within a rather
restricted range of the $(V,t_b')$ space. By increasing $V$ or $t_b'$ by a few
percent one reduces $T_c(SDW_0)$ below $T_c(SDW_{\pm})$ in the whole $B$ domain. On
the other hand by decreasing $V$ by a few percents one gets a hump in $T_c(SDW_0)$
on the left of the transition SDW$_0$ --  SDW$_{\pm}$.

The rapid oscillations in observable response functions are related to the oscillations of
$\delta$ \cite{Gorkov}, shown in  Fig.\ref{fd}(b). 
 Generally, we expect RO to be
visible if two conditions are fulfilled.
First, in order to have an overgap interference one needs a moderate MB parameter,
$\kappa \sim 1$. Second, one has to be in the oscillating regime on the scale $r$,
which is equivalent to $\kappa <\rho(V/t_b)\equiv 2(V/t_b)^{-2}\sqrt{(\gamma
V/t_b)^2+1}$. Fig.\ref{fd}(b) shows how the energy tilt
$\omega_c \delta$ varies in various parts of the phase diagram. At 30 Tesla we have
$\kappa\sim 0.5$ and $\rho(0.85)\approx3.3$, so that both conditions are fulfilled.
The effect is expected to be even stronger for higher fields because $\kappa$ then
increases.

 The maximal value of the critical temperature in Fig.\ref{fd},
$T_c^{\mbox{max}}\approx 1.1K$, is considerably smaller than the experimental value
of 5.5K. In this respect we note that $T_c^{\mbox{max}}$ is essentially model
dependent quantity, i. e. that the Hamiltonian (\ref{hkin}) represents a {\em
minimal} model for understanding the interplay between two SDW phases in the
magnetic field. Namely, recent experiments \cite{Lepeleven}
suggest that the anion ordering in TMTSF$_2$ClO$_4$ induces also, beside a strong
dimerizing potential $V$, rather large changes in other band parameters.

The present treatment also
does not include the quantitative analysis of the splitting of degeneracy of two
intraband phases, $SDW_+$ and $SDW_-$. Physically the degeneracy is lifted because
the realistic tight-binding dispersion along the chain is not strictly linear.
Consequently the dominant instability will be that of $SDW_-$, as discussed in
Ref.\cite{ZB01}. Similar conclusions were obtained also by numerical
calculations \cite{Kishigi98}, but without taking into account the two
component aspect of the order parameter (\ref{mag}). 
The second critical temperature can be calculated within Landau
theory as in Ref.\cite{Dupuis02}, and by taking the nonlinearity of the band
dispersion into account. The subphases of the high-field phase correspond to $SDW_+$
phases within $SDW_-$, each one nesting its own pair of Fermi sheets. Such scenario
is impossible for SDW$_0$ since it proceeds through nesting of all four sheets at
the {\em single} critical temperature.

In conclusion, we have solved exactly the one-particle problem of dimerized Q1D band
of electrons in magnetic field. Observables contain characteristic periodicity in
$1/B$, consistent with $260$ Tesla oscillations in normal and SDW phases of
(TMTSF)$_2$ClO$_4$. Using matrix RPA for SDW susceptibility we reproduce the
first-order transition between two types of FISDW ordering, as well as the overall
profile of the experimental phase diagram.




\begin{references}
\bibitem{review98} T. Ishiguro, K. Yamaji, and G. Saito, Organic
Superconductors IIe, (Springer-Verlag, Berlin, 1998).

\bibitem{Chaikin} P. M. Chaikin, J. Phys. I (France) 
{\bf 6}, 1875 (1996); P. Lederer {\em ibid.} {\bf 6}, 1899 (1996);
V. M. Yakovenko and H. S. Goan {\em ibid.} {\bf 6}, 1917 (1996).

\bibitem{Biskup99} N. Bi\v{s}kup {\em et al.}, Phys. Rev. B {\bf 60}, 15005
(1999). 



\bibitem{QNM} For historical references  and recent developments of the quantum
nesting model, see A. G. Lebed, Phys. Rev.  Lett. {\bf 88}, 177001 (2002).

\bibitem{Qualls00}J. S. Qualls {\em et al.}, Phys. Rev. B {\bf 62}, 12680 (2000). 

\bibitem{Matsunaga} N. Matsunaga  {\em et al.}, Phys. Rev. B {\bf 62}, 8611
(2000); cond-mat/0206010.

\bibitem{McKernan95}S. K. McKernan {\em et al.}, Phys. Rev. Lett. {\bf 75}, 
1630 (1995).

\bibitem{Chung00} O.-H. Chung {\em et al.}, Phys. Rev. B {\bf 61}, 11 649
(2000).

\bibitem{Kang01} W. Kang {\em et al.}, Synthetic Metals {\bf 120}, 1073 (2001).


\bibitem{Bjelis-Maki92} A. Bjeli\v{s} and K. Maki, Phys. Rev. B {\bf 45},
12887 (1992).

\bibitem{Uji97} S. Uji {\em et al.}, Solid State Commun.{\bf 103}, 387 (1997).

\bibitem{ZB01} D. Zanchi and A. Bjeli\v{s}, Europhys. Lett. {\bf 56}, 596
(2001).

\bibitem{Dupuis02}K. Sengupta and  N. Dupuis,  Phys. Rev. B {\bf  65}, 
035108 (2002).


\bibitem{Stark74} R. W. Stark and C. B. Friedberg, J. Low Temp. Phys. {\bf
14}, 111 (1974).

\bibitem{Osada92} T. Osada {\em et al.}, Phys. Rev. Lett. {\bf 69}, 1117 (1992).

\bibitem{Gorkov} L. P. Gor'kov and A. G. Lebed, 
Phys. Rev. B {\bf 51}, 3285 (1995); ibid. 1362.


\bibitem{Yamaji82} K. Yamaji, J. Phys. Soc. Jpn. {\bf 51}, 2787 (1982).

\bibitem{future} D. Radi\'c, A. Bjeli\v{s} and D. Zanchi, to be published.

\bibitem{eqdif} S. L. Ross, Diferential Equations, 3rd edition (John Wiley
\& Sons, New York, 1984). pp. 505-521; E. L. Ince, Ordinary Differential
Equations (Dover Publ., 1956) pp. 384,503,507; E. Kamke,
Differentialgleichungen (Akademische Verlag. Becker \& Erler Kom.-Ges.,
Leipzig, 1942).

\bibitem{Yoshino99} H. Yoshino {\em et al.}, J. Phys. Soc. Jpn. {\bf 68},
3142 (1999).

\bibitem{Lepeleven} 
 D. Le P\'{e}velen {\em et al.}, Eur. Phys. J. B{\bf 19}, 363 (2001).



\bibitem{Kishigi98} K. Kishigi, J. Phys. Soc. Jpn. {\bf 67},
3825 (1998).













\end{references}
\end{document}